\documentclass{elsarticle}
\usepackage{amssymb}
\usepackage{amsthm}
\usepackage{amsmath}

\begin{document}

\begin{frontmatter}

\title{On the definition of the domain growth-rate constant on a two-dimensional substrate
}

\author{Kazuhiko Seki}
\ead{k-seki@aist.go.jp}
\address{National Institute of Advanced Industrial Science and Technology (AIST)\\
Tsukuba Onogawa 16-1 AIST West, Ibaraki, 305-8569, Japan}




\begin{abstract}
In chemical vapor deposition (CVD) methods, the domains grow as a result of the attachment of diffusing 
surface bound species to islands formed by solid domains. 
The process of single-domain growth under two-dimensional diffusion is formulated by considering the movement of the domain boundary.
First, two definitions of the rate constant associated with the growth of the domain area are discussed: 
the first definition entails dividing 
 the domain size by the duration of the CVD growth, and the second dividing the area by time. 
 Then, the domain size is shown to be proportional to time for reaction-limited growth and 
 the domain area proportional to time for diffusion-limited growth. 
In addition, the growth rate of the domain area changes from that of reaction-limited growth to that of diffusion-limited growth as 
the domain size increases 
beyond a characteristic size.
\end{abstract}


\end{frontmatter}


\section{Introduction}
\label{sec:I}

The chemical vapor deposition (CVD) method has been widely used to synthesize high-quality large-area graphene layers with low defect densities.\cite{Li_09,Ago_10,Petrone_12,Chen_15,Seah_14,Tetlow_14,Feng_19}
In this regard, the CVD method has been widely used on Cu substrates, which yield monolayer graphene. 
\cite{Taira_2017,McCarty_09,Losurdo_11,Celebi_13,Jiang_18,Jaisi_18}
Recently, a millimeter-sized large-area monolayer of graphene was synthesized and its evolution observed in real time using radiation-mode optical microscopy (Rad-OM).\cite{Terasawa2015}
The shape of a growth domain is typically hexagonal, but other shapes, including circular shapes, can be formed. \cite{Taira_2017,McCarty_09,Losurdo_11,Celebi_13,Jiang_18,Jaisi_18}

The rate constant associated with the growth of a domain area can be defined in two ways. 
The first definition divides 
 the domain size by the duration of CVD growth, and the second divides the area by time. \cite{Kim_12,Terasawa2015,KATO_2016}
 The optimization of various factors controlling the growth process has been discussed using either or both of these definitions. 
However, quantitative studies of the growth of the domain area necessarily require an appropriate definition of the area growth rate. 
For example, the activation energy required for the incorporation of the 
surface-bound carbon species into the graphene domain under diffusion can be determined  
if the rate constant for domain area growth is properly defined using the incorporation  velocity of the 
surface-bound carbon species into the domain. 
In the first part of this study, the way in which the area growth rate is defined using the size or area and 
the relationship between the two above-mentioned definitions are discussed. 

The classical nucleation theory 
of Burton, Cabrera, and Frank (the BCF model) introduces the concept of 
radial velocity at the domain boundary.  
They showed this velocity to be independent of the radius when the diffusion length is smaller than the domain size. \cite{Burton_51}
As a consequence,  
the domain radius is proportional to time. 
In this study,  
the area is demonstrated to be proportional to time when the opposite limit is imposed, i.e., when the domain size is smaller than the diffusion length 
in the BCF model. 
When circular domain growth is studied with the BCF model, 
the interchange of molecules between the domain and the surface is assumed to be sufficiently rapid, such that  
the density of the surface-adsorbed species near the edge of the domain is maintained equal to that which would be in equilibrium. \cite{Burton_51}
Another assumption they made was that 
the influence of boundary movement on the diffusion problem was negligible. \cite{Burton_51}
In the study presented in this paper, these two points are addressed by taking into account 
the influence of the movement of the circular domain on the diffusion of surface-adsorbed species and by scrutinizing the boundary condition 
for the diffusion problem in the previous paper. \cite{Seki_19}

In the previous study, \cite{Seki_19}
the growth of the circular domain was investigated by taking into account 
the movement of this domain when solving the diffusion problem  
mainly by considering the limit of the rapid interchange of molecules between the domain and the surface. 
A phenomenological second order rate constant is introduced to study the 
reaction-limited growth, in a situation in which the interchange of molecules between the domain and the surface is slow. 
Here, the relationship between the second-order rate constant and 
the incorporation  velocity of the 
surface-bound species into the domain is closely examined.  
The second-order rate constant is shown to be proportional to the domain radius rather than remaining constant. 
As a result, the reaction-limited growth changes into diffusion-limited growth as 
the domain size increases; 
the domain growth-rate constant defined by dividing the size by time is obtained for reaction-limited growth and 
the domain growth-rate constant defined by dividing the area by time, is obtained for diffusion-limited growth. 
The conditions under which the reaction-limited growth and diffusion-controlled growth are divided 
are derived. 
The BCF model does not consider the slow interchange of molecules between the domain 
when circular domain growth is studied. \cite{Burton_51}
Therefore, the crossover from the initial reaction-limited domain growth to diffusion-limited growth is not accommodated in the BCF model for circular domains. 

In this work, it is shown that the division of the area by time is virtually independent of time in the BCF model 
when the domain size is smaller than the diffusion length. 
In the original BCF model, the size divided by the time was shown to be independent of the domain radius when the diffusion length was smaller than the domain size. \cite{Burton_51}
In this way, we validate the assumption of the BCF model that the influence of boundary movement on the diffusion is negligibly small.

\section{Definition of the domain-growth rate: size vs. area}
\label{sec:II}

For simplicity, the isotropic domain growth on the surface of the substrate is considered in terms of the increase in the radius $R(t)$. 
If the domain growth is limited by the attachment of surface-bound species rather than diffusion, 
the domain-growth rate is 
proportional to the peripheral length multiplied by 
the incorporation velocity of the 
surface-bound carbon species into the domain ($\kappa_{\rm i}$), \cite{Terasawa2015,Huijun_15}
\begin{align}
\frac{\partial \pi R^2(t)}{\partial t} \propto 2 \pi R(t) \kappa_{\rm i}.  
\label{eq:4}
\end{align}
The dimensionality of $\kappa_{\rm i}$ is [Length]/[Time]. 
If the incorporation of surface-bound carbon species into the domain is effectively given by the rate constant, as shown in Eq. ~ (\ref{eq:4}), 
then 
\begin{align}
R(t) \propto t .  
\label{eq:5}
\end{align}
Here and below, $R(t) \gg R(0)$ is assumed.
Therefore, the size is proportional to time for reaction-limited growth. 

The domain growth-rate constant defined by using the area follows if 
the growth rate of the domain area is 
proportional to the peripheral length multiplied by the concentration gradient at the domain boundary \cite{Seki_19}
\begin{align}
\frac{\partial \pi R^2(t)}{\partial t} \propto \left. 2 \pi R(t) D\frac{\partial C}{\partial r} \right|_{r=R(t)}, 
\label{eq:1}
\end{align}
where $D$ and $C$ denote the diffusion constant and concentration of  surface-bound carbon species, respectively. 
When the concentration of the surface-bound carbon species obeys the Laplace equation (diffusion equation) in two dimensions, we have: 
\begin{align}
\frac{\partial C}{\partial r}\propto \frac{1}{r}.
\label{eq:2}
\end{align}
By substituting Eq. (\ref{eq:2}) into Eq. (\ref{eq:1}), we find that \cite{Kim_12,Seki_19}
\begin{align}
R^2(t) \propto t . 
\label{eq:3}
\end{align}
Therefore, the domain area is proportional to time for diffusion-limited growth. 

Here, the boundary condition around the circular domain can be introduced using 
the second-order rate constant in two dimensions denoted by $k_{\rm 2D}$ as \cite{Seki_19}
\begin{align}
 \left. 2 \pi R(t) D\frac{\partial C}{\partial r} \right|_{r=R(t)}=k_{\rm 2D} \left[C(R(t),t)-C_{\rm e} \right], 
\label{eq:6_2}
\end{align}
where $k_{\rm 2D}$ should have the same dimensionality as $D$, which can be expressed as [Length]$^2$/[Time].
The second-order rate constant $k_{\rm 2D}$ is proportional to the incorporation frequency $\nu_i$ times $2 \pi R(t) dr$, 
where $\nu_i$ has the dimension of inverse time, and $dr$ denotes the infinitesimal radial component around the periphery of the domain. 
Therefore, $k_{\rm 2D}$ scales with $R(t)$. 
To reflect this scaling in the boundary condition, 
it is more appropriate to express 
the boundary condition around the circular domain using the incorporation  velocity 
of the 
surface-bound species into the domain denoted by $\kappa_{\rm i}$ as 
\begin{align}
 \left. D\frac{\partial C}{\partial r} \right|_{r=R(t)}=c_{\rm g} \kappa_{\rm i} \left[C(R(t),t)-C_{\rm e} \right],
\label{eq:6}
\end{align}
where $C_{\rm e}$ represents the equilibrium concentration at the periphery of the circular domain. 
At equilibrium, the incorporation and detachment of surface bound carbon species are balanced. 
Further, $c_{\rm g}$ is a geometrical constant that is independent of the domain radius.  
The value of this constant, $c_{\rm g}=1/2$ is obtained for circular domains because 
half of the surface-bound species in the shell of the area $2 \pi R(t) dr$ on the domain boundary moves inward in the radial direction. 

Previously, in chapter 6 of Ref. \cite{Burton_51}, 
the rate of advance of a circular domain was considered. 
In the BCF model, $C(R(t),t)=C_{\rm e}$ is assumed by considering the limit of the fast exchange of adsorbed species,  $\kappa_{\rm i} \rightarrow \infty$;  
the interchange of molecules between the domain and the surface is sufficiently rapid  
such that $C(R(t),t)\approx C_{\rm e}$ holds.  
The probability current of surface-bound species flowing into the domain from the surface is obtained as: 
$j_<= \left. 2 \pi R D \partial C/(\partial r) \right|_{r=R}$. 
A similar probability current can be defined for atoms in the domain, $j_>=- \left. 2 \pi R D \partial C_{\rm D}/(\partial r) \right|_{r=R}$, 
where $C_{\rm D}$ indicates the concentration of atoms on the surface of the domain; in addition, 
the minus sign indicates that the flow direction is opposite for atoms in the domain in that they are flowing outward at the domain edge. 
When the diffusion of $C_{\rm D}$ is described by the same equation as that of $C$, $j= j_< + j_>$ is obtained as 
$j=2\pi D (C_\infty-C_{\rm e}) /\left[ I_0 (R/x_{\rm D}) K_0(R/x_{\rm D})\right]$, \cite{Burton_51},
where 
$I_n(z)$ and $K_n(z)$ are the modified Bessel functions of the first and second kinds, respectively, \cite{Abramowitz}
$C_\infty$ indicates the surface concentration of the surface-bound species sufficiently far away from the domain, 
$x_{\rm D}$ is the diffusion length given by $\sqrt{D/k_d}$, and $k_d$ is the desorption rate per unit area of the surface-bound species to the gas phase. 
By considering the mass conservation at the periphery of the domain, 
we have $\rho 2 \pi R dR/(dt)= j$, where $\rho$ is the two-dimensional atomic density of the domain. 
(Note that $\rho 2 \pi R dR/(dt)= j$ is not used to solve the diffusion problem in the BCF model.
The solution of the diffusion problem is expressed using $I_0 (R/x_{\rm D})$ and $K_0(R/x_{\rm D})$, for which 
the boundary condition is $C(R(t),t)=C_{\rm e}$.)
When the diffusion length exceeds the domain size denoted by $R$, 
we can introduce $I_0(z) \approx 1$ and $K_0(z) \approx \ln(2/z)-\gamma$,  
where $\gamma=0.577 \cdots$ denotes Euler's constant.  \cite{Abramowitz}
In this limit, we obtain 
\begin{align}
R \frac{d R(t)}{dt} 
&\approx \frac{D}{\ln\left(2 x_{\rm D}/R\right)-\gamma}
\frac{(C_\infty-C_{\rm e})}{\rho}. 
\label{BCF_r}
\end{align}
The same limit can be obtained even when $j\approx j_<$. 
The contribution of the probability current for atoms in the domain can be ignored when the diffusion length exceeds the domain size. 
By ignoring the weak logarithmic correction term, the solution to Eq. (\ref{BCF_r}) can be written as 
\begin{align}
R^2(t)&\approx
\frac{2 Dt}{\ln\left(2 x_{\rm D}/R\right)-\gamma}
\frac{(C_\infty-C_{\rm e})}{\rho}. 
\label{BCF_r1}
\end{align}
In the BCF model, \cite{Burton_51}
$dR/(dt)=2 x_{\rm D}(C_\infty-C_{\rm e})/\rho$ was derived by considering 
the opposite limit of $R > x_{\rm D}$, 
where $j\approx j_<$ should not be taken instead of $j= j_< + j_>$. (Equation (32) in Ref. \cite{Burton_51}.)
To summarize the results of the BCF model, 
the area is proportional to the time when $R< x_{\rm D}$, whereas the size is proportional to the time when $R> x_{\rm D}$ 
in the limit of the rapid interchange of molecules between the domain and surface. 
The rapid interchange limit can be expressed by the limit $\kappa_{\rm i} \rightarrow \infty$ of the incorporation velocity.  
Below, the case in which the interchange of molecules between the domain and the surface is finite is presented. 
It is shown that 
the area is proportional to the time when the domain size exceeds the critical radius given by $4D/\kappa_{\rm i}$, apart from the logarithmic correction term 
when $R< x_{\rm D}$; furthermore,  
the domain size is proportional to time
when the domain size is smaller than the critical radius.  
In addition to considering the slow interchange of molecules between the domain and the surface, 
the movement of the periphery of the domain is also considered in the diffusion problem.
Here, $j= j_< + j_>$ is approximated as $j\approx j_<$ because the limit of $R < x_{\rm D}$ is considered. 
In the BCF model, the Gibbs-Thomson effect was considered by 
$C_{\rm e}= C_0 \exp(a/R)$,  
 where $C_0$ is the equilibrium concentration at the periphery of the flat interface, and $a$ is a positive constant. 
 In this work, the Gibbs--Thomson effect is ignored for simplicity. 

To obtain the growth rate, the boundary condition is supplemented by 
the condition describing the effect of the movement of the periphery of the domain 
on the diffusion problem. \cite{Krapivsky_12,Larralde_93,Burlatsky_96,Oshanin_98,Seki_19}
For isotropic growth, the condition is obtained from the mass conservation law  \cite{Seki_19}
\begin{align}
2\pi R(t)\left(\rho- C(R(t),t)- k_d \int_0^t d t_1 C(R(t),t_1) +gt\right)\frac{\partial R}{\partial t}
=\left. 2\pi D R(t) \frac{\partial C}{\partial r} \right|_{r=R(t)}, 
\label{eq: consdiffo}
\end{align}
where $\rho$ is the two-dimensional atomic density of the graphene domain,
$k_d$ is the desorption rate per unit area of the surface-bound carbon species to the gas phase, and 
$g$ is the rate at which the carbon species is deposited on the substrate surface from the gas phase per unit area. 
The left-hand side of Eq. (\ref{eq: consdiffo}) indicates the mass change in the growing interface area 
given by $2\pi R(t) d R$ during $t$ and $t+dt$ and the right-hand side indicates the mass current flowing into the interface from the two dimensional surface. 
We assume that the dominant contribution 
originates from $\rho$ and $C(R(t),t)$ and obtain  \cite{Seki_19}
\begin{align}
\left[\rho- C(R(t),t) \right] \frac{\partial R}{\partial t}
=\left. D \frac{\partial C}{\partial r} \right|_{r=R(t)}. 
\label{eq:bcC1}
\end{align}
The influence of the movement of the boundary on the diffusion problem is given by Eq. (\ref{eq:bcC1}). 
The classical BCF model neglects the influence of boundary movement on the diffusion problem.  
Equation (\ref{eq:bcC1}) can be regarded as the Stefan boundary condition for describing the boundary motion. \cite{crank_87,Gupta_18}
The boundary conditions given by Eqs. (\ref{eq:6}) and the Stefan boundary condition given by Eq. (\ref{eq:bcC1}) 
were used to solve the diffusion equation of surface-bound species. \cite{Seki_19}

When the diffusion length defined by $x_{\rm D}=\sqrt{D/k_d}$ exceeds the domain size, denoted by $R$, 
the approximate solution can be expressed as \cite{Seki_19}
\begin{align}
R^2(t)
&\approx 
\frac{2 Dt}{\ln\left(2 x_{\rm D}/R\right)-\gamma+2D/(\kappa_{\rm i}  R)} \frac{(C_\infty-C_{\rm e})}{\rho}, 
\label{eq:7}
\end{align}
where 
$C_\infty$ indicates the surface concentration of the surface-bound species sufficiently far away from the domain,  
$\gamma=0.577 \cdots$ is Euler's constant, and 
$k_{\rm 2D}=\pi R(t) \kappa_{\rm i}$ is introduced, 
where half of the surface-bound species in the shell of the area $2 \pi R(t) dr$ on the domain boundary is assumed to move inward in the radial direction, 
as explained below Eq. (\ref{eq:6}). 
The relation $k_{\rm 2D}=\pi R(t) \kappa_{\rm i}$, which has not been recognized previously, \cite{Seki_19} 
enables us to define the critical radius that divides diffusion- and reaction-limited growth. 

For the diffusion-limited growth, 
\begin{align}
R \left( \ln\left(2 x_{\rm D}/R\right)-\gamma \right)> \frac{2D}{\kappa_{\rm i}}. 
\label{eq:df}
\end{align}
The approximate growth rate law is obtained from Eq. (\ref{eq:7}) as  Eq. (\ref{BCF_r1}). \cite{Seki_19}
For diffusion-limited growth, $R^2(t)/t$ provides the growth rate constant, except for the weak logarithmic correction term. 
The BCF theory neglects 
the motion of the domain boundary when attempting to solve the diffusion problem. 
Because a consistent result is obtained here, this assumption is justified. 

For reaction-limited growth, 
\begin{align}
R \left(\ln\left(2 x_{\rm D}/R\right)-\gamma \right) <\frac{2D}{\kappa_{\rm i}} ,  
\label{eq:rl}
\end{align}
the approximate growth rate law is obtained as  
\begin{align}
R(t) &\approx \kappa_{\rm i}\frac{(C_\infty-C_{\rm e})}{\rho} t,
\label{eq:expression4}
\end{align}
with $R(t)/t$ providing the growth rate constant. 
Equation (\ref{eq:expression4}) can be expressed as 
the mass conservation law at the periphery of the domain, 
$\rho v_n \approx \kappa_{\rm i}(C_\infty-C_{\rm e})$, where the radial domain-growth rate is given by $v_n=R(t)/t$.  
In addition, $\rho v_n$ indicates the rate at which the size of the solid domain increases with density $\rho$, which is equal to $\kappa_{\rm i}(C_\infty-C_{\rm e})$, which
indicates the rate at which surface bound species are incorporated.

Therefore, we find that $R(t)$ is proportional to time for reaction-limited growth and 
$R^2(t)$ is proportional to time for diffusion-limited growth. 
Interestingly, the characteristic size can be defined to distinguish between reaction- and 
diffusion-limited growth. 
The characteristic size can be obtained by solving 
\begin{align}
R_c \left( \ln\left(2 x_{\rm D}/R_c\right)-\gamma\right) \approx \frac{2D}{\kappa_{\rm i}} . 
\label{eq:rc}
\end{align}
If the domain is smaller than $R_c$, the growth is reaction limited; 
if the domain is larger than $R_c$, the growth is diffusion limited. 
In other words, 
the domain growth can initially be reaction limited and eventually become diffusion limited. 

The approximate solution to extrapolate the initial reaction-limited growth and the subsequent diffusion-limited growth can be 
obtained by substituting $R_c$ for $R$ in $\ln[4D/(k_d R^2)]$ on the right-hand side of Eq. (\ref{eq:7}). 
For convenience, we define the degree of supersaturation of the two-dimensional surface as $\sigma=(C_\infty-C_{\rm e})/C_{\rm e}$. 
The result can be expressed using the dimensionless domain size denoted by 
$\bar{R}=R\kappa_{\rm i}/(2D)$ and the dimensionless time denoted by 
$\bar{t}=t \left(C_{\rm e} \sigma/\rho \right)\left(\kappa_{\rm i}^2/D\right)$ as
\begin{align}
\bar{R}(t) &\approx 
\frac{-1+
\left\{1+
\bar{t} \left[\ln \left(C_{\rm iD}^2/\bar{R}_c^2 \right)
-2\gamma
\right] 
 \right\}^{1/2}}
{\ln \left(C_{\rm iD}^2/\bar{R}_c^2 \right)-2\gamma}  
\label{eq:8_0}\\
&\approx\frac{-1+\left\{1+ \bar{t}\ln \left(C_{\rm iD}^2/\bar{R}_c^2 \right)\right\}^{1/2}}{\ln \left(C_{\rm iD}^2/\bar{R}_c^2 \right)},  
\label{eq:8}
\end{align}
where 
the dimensionless characteristic size is defined by $\bar{R}_c=R_c\kappa_{\rm i}/(2D)$ and 
$C_{\rm iD}=\kappa_{\rm i}/\sqrt{Dk_d}$. 
Here, $C_{\rm iD}$ can be rewritten as $\left(\kappa_{\rm i}/k_d\right)/\sqrt{D/k_d}$; that is, 
the incorporation velocity multiplied by the lifetime $1/k_d$ of the surface-bound species is divided by the diffusion length. 

\begin{figure}
\centerline{
\includegraphics[width=0.6\columnwidth]{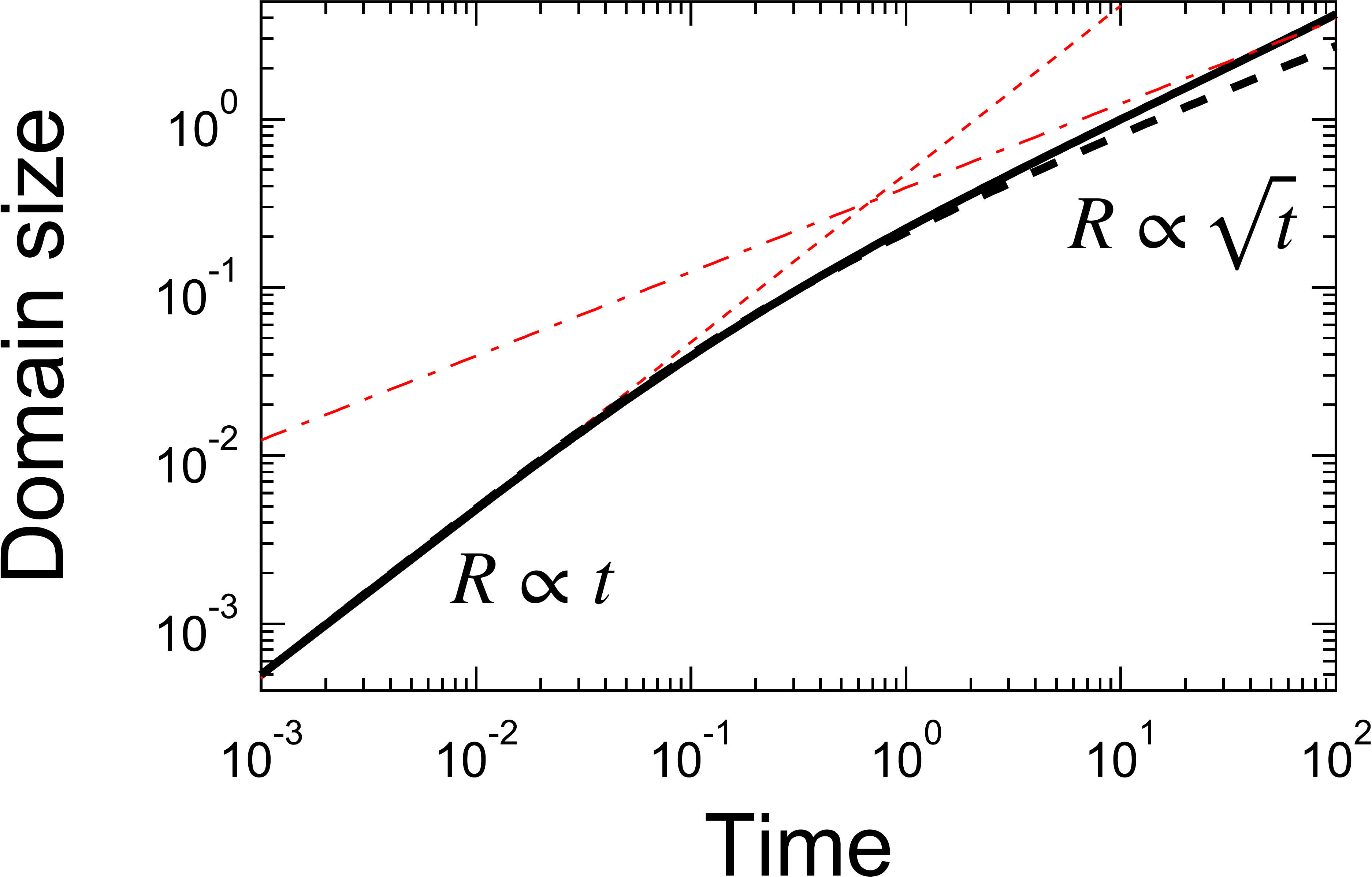}
}
\caption{(Color online) Dimensionless domain size defined by 
$R\kappa_{\rm i}/(2D)$ plotted against the dimensionless time defined by $t \left(C_{\rm e} \sigma/\rho \right)\left(\kappa_{\rm i}^2/D\right)$, 
where $C_{\rm iD}=\kappa_{\rm i}/\sqrt{Dk_d}=10^2$. 
The thick solid line was obtained by numerically solving Eq. (\ref{eq:7}). 
The thick dashed line indicates the approximate solution given by Eq. (\ref{eq:8}), where $\bar{R}_c=0.173$.
The (red) thin dashed line indicates the $R \propto t$ relation. The (red) dashed-dotted line indicates the $R \propto \sqrt{t}$ relation. 
}
\label{fig:1}
\end{figure}

In Fig. \ref{fig:1}, we show the numerically exact result obtained from Eq. (\ref{eq:7}) using Newton's method, 
where the seed for the method is given by the approximate result obtained from Eq. (\ref{eq:8}). 
The domain-growth curves shown in Fig. \ref{fig:1} are characterized by a parameter given by $C_{\rm iD}$. 
The larger the value of $C_{\rm iD}$, the clearer the transition between reaction- and diffusion-limited growth. 
We obtain $\bar{R}_c=0.173$ for $C_{\rm iD}=10^2$. 
When the size is smaller than $\bar{R}_c=0.173$, 
the domain size is proportional to time as predicted from reaction-limited growth. 
When the size is larger than $\bar{R}_c=0.173$, 
the domain size is proportional to $\sqrt{t}$, as predicted from diffusion-limited growth. 
The qualitative feature was reproduced using Eq. (\ref{eq:8}), although the domain size is underestimated. 

The above results are obtained when the diffusion length, defined by $x_{\rm D}$, exceeds the domain size denoted by $R$.  
By assuming a two-dimensional diffusion constant of $10^{-6}$ m$^2$s$^{-1}$ and a size of $10 \sim 20 \mu$m, \cite{Terasawa2015,Kim_12,KATO_2016}  
the mean lifetime of the adsorbed species before being evaporated again should be larger than $2.25\sim 4 \times 10^{-4}$ s, which is 
estimated from $x_{\rm D}>R$, $x_{\rm D}=\sqrt{D /k_d}$, and the mean lifetime is given by $1/k_d$. 
The mean lifetime may depend on the gas flow rate, temperature, and substrate, and is difficult to measure. 
The domain growth law in the opposite limit of $x_{\rm D}<R$ is explicitly given in the BCF model. \cite{Burton_51}
It should be pointed out that the domain growth law depends on the relative magnitudes of the diffusion length and domain size. 
These theoretical results on an isolated closed domain could be useful for interpreting the measured domain-growth rates of graphene 
on a Cu substrate. \cite{Terasawa2015,Kim_12,KATO_2016}

In Fig. \ref{fig:1}, the theoretical prediction of the crossover from reaction- to diffusion-limited growth is plotted and the experimental conditions under which the crossover would be observed are predicted. 
The critical domain size could be roughly estimated from $ R_c \approx 4 D/\kappa_{\rm i}$. 
To observe the crossover at approximately $10 \mu$m, the incorporation velocity would be expected to be approximately $0.4$ m s$^{-1}$. 
The growth law below $R_c$ can be expressed as  
$R(t)/t \approx \kappa_{\rm i}C_{\rm e} \sigma/\rho$. 
To be consistent with the observed value of $R(t)/t \approx 0.02\sim 0.1 \mu$m  s$^{-1}$,  \cite{Terasawa2015}
$C_{\rm e} \sigma/\rho=0.25 \times 10^{-6}$ is estimated using 
$\kappa_{\rm i}=0.4$ m s$^{-1}$. 
The incorporation velocity would be required to be lower to observe the crossover at a higher value of $C_{\rm e} \sigma/\rho$. 
The incorporation velocity might depend on the temperature and flow rate of H$_2$. \cite{Dong_19}
A possible approach that could be used to confirm the crossover would be to vary these parameters.  

It should be noted that, when experiments are conducted, multiple domains are growing at any time, and these domains could interfere with 
each other in the diffusion-limited regime. \cite{Terasawa2015}
If the area is proportional to time, as predicted for diffusion-limited growth within a certain time, and the growth is subsequently suppressed,  
interference by the diffusion might be the reason for the suppression. 
In this case, 
the crossover could be examined using the data of the earlier growth period by determining 
whether the size is proportional to time before the area becomes proportional to time.

\section{Conclusion}
\label{sec:III}
The growth rate of an isotropic domain on a substrate surface was studied by considering the attachment and detachment kinetics at the domain boundary of 
 surface diffusing carbon species. 
The desorption rate per unit area of the surface-bound carbon species in the gas phase and 
the deposition rate of carbon species from the gas phase to the substrate surface per unit area were also taken into account.  
The concentration of surface-bound species far away from the center of the domain becomes constant and is determined by the 
ratio between the deposition and desorption rates. 

Approximate analytical expressions for the domain area growth rate are obtained; the growth rate of the domain area is expressed as a function of the two-dimensional diffusion constant, 
the incorporation velocity of surface-bound species into the solid domain, the degree of supersaturation of the two-dimensional surface, 
and the lifetime of the surface-bound species, given by the inverse of the desorption rate constant. 
 Our study of the previously obtained solution of the corresponding Stefan problem \cite{Seki_19}
showed that the size is proportional to time for reaction-limited growth, and the domain area is proportional to time for diffusion-limited growth. 
Here, it was also shown that the initial domain growth is limited by the reaction at the domain boundary, and transforms into diffusion-limited growth in the subsequent time regime when the domain size exceeds the characteristic size. 

An analytical expression for diffusion-limited growth was also derived from the relation between the radial velocity and the domain radius expressed using the modified Bessel functions 
in the BCF model. \cite{Burton_51}
The effect of the movement of the periphery of the domain on the diffusion problem is disregarded in the BCF model, 
whereas it is taken into account by solving the Stefan problem. \cite{Seki_19}
The assumption in the BCF model regarding the effect of the movement of the periphery of the domain on the diffusion problem is justified. 
The BCF model assumes that, when circular domain growth occurs, the interchange of molecules between the domain and the surface is assumed to be rapid. 
Therefore, the crossover from the initial reaction-limited domain growth to diffusion-limited growth is not accommodated in the BCF model for circular domains. \cite{Burton_51}
Importantly, these results are obtained when the domain radius is smaller than the diffusion length. 
For the opposite limit, i.e., when the diffusion length is shorter compared with the domain radius, 
the radial velocity is shown to be independent of the domain radius, and its size is proportional to time in the absence of the Gibbs-Thomson effect. \cite{Burton_51}
The comparative theoretical results presented here provide a foundation for the study of the factors that control domain growth. 

\section{Acknowledgments}
I would like to thank the anonymous reviewers for their insightful comments on the BCF model.




\end{document}